\documentclass[final,3p,times]{elsarticle}
\usepackage{soul}
\usepackage{amsmath}
\usepackage{tablefootnote}
\usepackage[colorlinks=true, allcolors=teal]{hyperref}

\usepackage{amssymb}
\usepackage{lipsum}
\usepackage{lineno}

\journal{Cryogenics}

\begin{document}
\begin{frontmatter}



\title{Superconducting properties of commercially available solders for low-field applications}

\author[NCSU,TUNL]{C.~Hickman}
\author[Montclair,Duke]{K.~K.~H.~Leung\corref{cor1}}
\ead{leungk@montclair.edu}
\cortext[cor1]{Corresponding author}
\author[NCSU]{A.~H.~Al-Tawhid}
\author[Caltech]{B.~W.~Filippone}
\author[NCSU,TUNL]{P.~R.~Huffman}
\author[NCSU,TUNL,NCSUNucEng]{E.~Korobkina}
\author[NCSU,Duke]{D.~P.~Kumah}
\author[Caltech,HRL]{C.~M.~Swank}

\affiliation[NCSU]{organization={North Carolina State University},
             addressline={Department of Physics},
             city={Raleigh},
             postcode={27695},
             state={NC},
             country={USA}}
\affiliation[TUNL]{organization={Triangle Universities Nuclear Laboratory},
             addressline={Department of Physics},
             city={Durham},
             postcode={27708},
             state={NC},
             country={USA}}
\affiliation[Montclair]{organization={Montclair State University},
             city={Montclair},
             postcode={07043},
             state={NJ},
             country={USA}}
\affiliation[Duke]{organization={Duke University},
             addressline={Department of Physics},
             city={Durham},
             postcode={27517},
             state={NC},
             country={USA}}
\affiliation[Caltech]{organization={California Institute of Technology,},
             addressline={Kellogg Radiation Laboratory},
             city={Pasadena},
             postcode={91125},
             state={CA},
             country={USA}}
\affiliation[NCSUNucEng]{organization={North Carolina State University},
             addressline={Department of Nuclear Engineering},
             city={Raleigh},
             postcode={27695},
             state={NC},
             country={USA}}
\affiliation[HRL]{organization={HRL Laboratories, LLC},
             addressline={3011 Malibu Canyon Road},
             city={Malibu},
             postcode={90265},
             state={CA},
             country={USA}}             
             
\begin{abstract}
 Solders with superconducting properties around $4\,{\rm K}$ are useful in low magnetic field environments for AC current leads or in electrical and mechanical bonds. Accurate knowledge of these properties are needed in high precision experiments. We have measured the electrical resistance of five commercially-available solders: 50\%Sn-50\%Pb, 60\%Sn-40\%Pb, 60\%Sn-40\%Pb-0.3\%Sb, 52\%In-48\%Sn, and 96.5\%Sn-3.5\%Ag, down to $2.3\,{\rm K}$ and in applied magnetic fields from 0 to 0.1$\,{\rm T}$. Their critical temperatures $T_c$ and critical fields $B_c$ were extracted in our analysis, taking into account the observed 90\%-to-10\% transition widths. Our best candidate for low-loss AC current leads in low fields is 50\%Sn-50\%Pb, which had zero-field $T_{c,0} = (7.1 \pm 0.3)\,{\rm K}$, and remained high to $T_c(B=0.1\,{\rm T}) = (6.9 \pm 0.3) \,{\rm K}$. We report $T_c$ and $B_c$ of 60\%Sn-40\%Pb-0.3\%Sb and $B_{c,0}$ of 96.5\%Sn-3.5\%Ag for the first time. Our $T_{c,0}= (3.31 \pm 0.04)\,{\rm K}$ for 96.5\%Sn-3.5\%Ag disagrees with a widely adopted value.
\end{abstract}



\end{frontmatter}

\section{Introduction} \label{sec:intro}

Many solders -- especially those with major Sn, Pb, Bi, or In components -- have superconducting transitions around or above normal boiling liquid helium temperatures. Solders with superconducting properties can be used for electrical-current transport in cryogenic environments where loss and heat avoidance is of importance. For AC current, solders may prove more desirable than "hard" superconducting wires (often Nb-based superconducting filaments embedded in a normal-conducting matrix) that have losses due to coupling and eddy current effects \cite{wilson_superconducting_1983, lyly_design_2013, wilson_nbti_2008,  norris_calculation_1970}. In sensitive, low-temperature, weak magnetic field applications, such as in SQUID-based NMR \cite{mcdermott_microtesla_2004, Barskiy2025} and quantum computing \cite{seedhouse_quantum_2021, brandl_cryogenic_2016}, unaccounted for superconducting transitions can lead to spurious effects and degradation of performance due to changes in thermal conductivity or distortion of magnetic fields.

The superconducting solder wires we studied consist of eutectic mixtures of type-I superconductors. They are easy to obtain and can be implemented as current leads. Our particular application is for a dilution-refrigerator-based nuclear physics experiment that will search for a time-reversal-symmetry-violating neutron electric dipole moment (nEDM) using novel NMR techniques on polarized free neutrons and $^3$He \cite{ahmed_new_2019, leung_neutron_2019, chupp_electric_2019, abel_measurement_2020}. For our $\sim 1\,{\rm \mu T}$ static field, the volume-averaged gradient specification is $< 50 \, {\rm pT \, cm^{-1}}$ over a $\sim 1\,{\rm m^3}$ volume. For spin dressing \cite{cohen-tannoudji_absorption_1969}, we also require a $\sim 1\,{\rm kHz}$ AC field with $\sim 0.1\,{\rm mT}$ amplitude. The designed $\cos(\theta)$ AC magnets needed to achieve the desired field uniformity require a $\sim 6-10\,{\rm A}$ current (amplitude). Due to engineering constraints, we are not able to submerge our magnet wiring in liquid helium. For maintaining below $T_c$ during operation,  one system \cite{ahmed_new_2019} will use exchange gas cooling with surfaces cooled by a helium circulation system; in another system \cite{cianciolo_systematics_2025} will rely on conductive cooling with a $4.2\,{\rm K}$ helium bath.

For our AC magnet wiring candidates, we chose solders made from an eutectic mixture of type-I superconductors \cite{murakami_anomalous_2023, arima_observation_2024} that can be manufactured as a single, relatively large (0.5-1.0$\,{\rm mm}$) diameter superconducting filament. This avoids the need for a normal-conducting matrix, and the only AC-related heating is a frequency-independent hysteretic loss \cite{bean_magnetization_1964}. We also want a high $T_c$ in our magnetic field range of interest of $\lesssim 50\,{\rm mT}$. Our baseline candidate was pure 100\% Pb, which has $T_{c,0} = 7.2\,{\rm K}$ and $B_{c,0} = 0.08\,{\rm T}$ \cite{daunt_lxx_1939}. We use $T_{c,0}$ and $B_{c,0}$ to denote $T_c(B=0)$ and $B_{c}(T=0)$, respectively.

The electric transport properties of four commercially available solders with $T_{c,0} \gtrsim 5.5\,{\rm K}$ were measured. Due to our interest in low fields, we did not exceed $100\,{\rm mT}$ applied field. We also incorporated the 90\%-to-10\% resistance transition width in our analysis and propagated the impact of this on our extracted parameters. This is important for the detailed deployment of these materials. For the most promising material for our AC field coil wiring, 50\%Sn-50\%Pb, which maintains a relatively high $T_c$, we describe a technique to apply a cryogenically robust electrical insulation to the solder wire. 

Many of the solders we characterized, such as the Sn-Pb solders, are widely used in cryostat construction \cite{ekin_experimental_2006,Pobell1996,White2002}. The Sb-doped Pb-Sn solder has been recommended because the Sb content helps to inhibit embrittlement and cracking from cryogenic thermal cycling \cite{ekin_experimental_2006}. And the silver-bearing tin solder, 96.5\%Sn-3.5\%Ag, is recommended for use in ``most mechanical joints'' \cite{ekin_experimental_2006} due to its balance between mechanical strength and low melting temperature ($221\,{\rm ^{\circ}C}$). We observe disagreements between our results and the references for 50\%Sn-50\%Pb, 52\%In-48\%Sn, and 96.5\%Sn-3.5\%Ag. And we are the first to report on the $T_c$ and $B_c$ of 60\%Sn-40\%Pb-0.3\%Sb and $B_{c,0}$ of 96.5\%Sn-3.5\%Ag.

\section{Methods}
\label{sec:experiment}
\subsection{Wire Specifications}
The 50\%Sn-50\%Pb solder wire has a diameter of 0.81$\,{\rm mm}$ and is from Prince \& Izant Co.~Inc. The 60\%Sn-40\%Pb wire used is a generic solder with a $\sim 1.6\,{\rm mm}$ diameter. The 60\%Sn-40\%Pb-0.35\%Sb is a 0.5$\,{\rm mm}$ diameter rosin core wire solder from Micro-Measurements (MMF006619-ND). The 52\%In-48\%Sn is a 0.8$\,{\rm mm}$ diameter solder from Chip Quik Inc. (SMDIN52SN48). The 96.5\%Sn-3.5\%Ag solder is from Harris\textregistered ~(Stay Brite SB31) with 1.6$\,{\rm mm}$ diameter, which has a specified melting temperature of 221$^\circ$C.

\subsection{Experimental Setup}
Our experiments were carried out inside a Quantum Design Physical Property Measurement System \textregistered \footnote{From the Department of Materials Science and Engineering, NC State University}, which can provide fields up to 9~T and a temperature range of 1.9 - 350~K. We used a standard 4-probe configuration to measure the DC resistance of the samples. A Keithley 6221 Ultra-sensitive Current Source, providing a 100$\,{\rm mA}$ amplitude and 5$\,{\rm Hz}$ square wave current, was used with a Keithley 2182A nanovoltmeter. This combination has a specified resistance measurement range of $10\,{\rm n\Omega}$ to $200\,{\rm M\Omega}$. Due to our interest in higher temperatures and low fields, we did not exceed 0.1$\,{\rm T}$ or cool below 2.3$\,{\rm K}$. The specified temperature stability and accuracy are $\pm 0.2\%$ and $\pm 1\%$, respectively, and the specified field uniformity is $\pm 0.01\%$ over the 5.5$\,{\rm cm}$ diameter region of the sample puck. The magnetic field was applied perpendicular to the current direction.

\begin{figure}[]
\centering 
\includegraphics[width=7 cm]{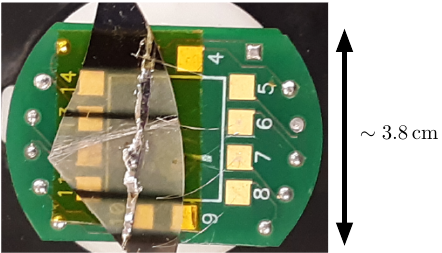}
\caption{A solder sample deposited on a sapphire substrate installed on top of the Quantum Design instrument's DC resistivity measurement puck.
\label{fig:1}}
\end{figure}

Our solder samples were melted and deposited on a single-crystal sapphire substrate (see Figure~\ref{fig:1}). Four aluminum wires were bonded to the sapphire and gold electrical contact pads of the measurement puck's printed circuit board.  The typical resistances measured at room temperature of our samples were between 0.05 to 1.5$\,{\rm m\Omega}$. Due to its irregular shape, we can only give a rough estimate for the average cross-section of the deposited solder as $\sim 4\,{\rm mm}^2$.

%Applied fields at 25, 50, 75, and 100 mT for higher Tc solders.
%staybrite went from 2.3 to 4 K. Applied fields at 5 and 10 mT
%I have InSn & 5050 ~ 1.5 mOhm, Ostalloy & 6040 ~0.1 mOhm, and StayBrite (Fig 2, 0.05 mOhm)

The self-field correction in these measurements due to the 100$\,{\rm mA}$ drive current is negligible. From Ampere's law, for a round conductor with radius $r$ and carrying a uniform current $I$, the largest self-field occurs at its surface \cite{wilson_superconducting_1983} and has a value of:
\begin{equation}
    B = \frac{\mu_0 I}{2 \pi r}.
    \label{eq:self_field}
\end{equation}
For the drive current used, and assuming $r\approx 0.5\,{\rm mm}$, the self-field correction is $< 0.1\,{\rm mT}$. This is small compared to the applied fields used in our $B_c$ measurements. It is also small compared to the fields involved in our operating magnet, where the local field at the wire locations will be $\lesssim 40\,{\rm mT}$.

\section{Results}\label{sec:Results}

We measured the resistance versus temperature curves in $\sim 0.1\,{\rm K}$ steps at different fields to observe the superconducting transition characterized by a drop to nominally zero resistance. We define the critical temperature using the resistive method \cite{warren_superconductivity_1969, IEC2005}, by which $T_c$ is determined when the resistance is 50\% of the normal state. Following IEC superconductivity standards, the width of the transition is determined as the range covering 10\% to 90\% of the normal resistance. For this paper, we used half of the transition width as the ($\pm$) uncertainty in the $T_c$ values reported.

With our $T_c$ measurements at different applied fields, we used the Werthamer-Helfand-Hohenberg (WHH) equation \cite{werthamer_temperature_1966},
\begin{equation}
    B_c(T) = B_{c,0}\left[1 – \frac{T^2}{{T_{c,0}}^2}\right],
    \label{eq:WHH}
\end{equation}
to determine $B_{c,0}$. Fitting was performed using a Levenberg–Marquardt non-linear $\chi^2$ algorithm. The $1\,{\rm \sigma}$ uncertainties inputted into the fit were $\pm \,(90\%-10\%  \rm{\,normal ~resistance})/2$ (described earlier) and the uncertainties in the fit parameters quoted below are $\pm 1\,{\rm \sigma}$ determined by the curvature in $\chi^2$ space. Two free parameters are used for the fit, $B_{c,0}$ and $T_{c,0}$, with the latter well anchored to the measured points at zero field. 

\begin{figure}[]
\centering 
\includegraphics[width=0.75\textwidth]{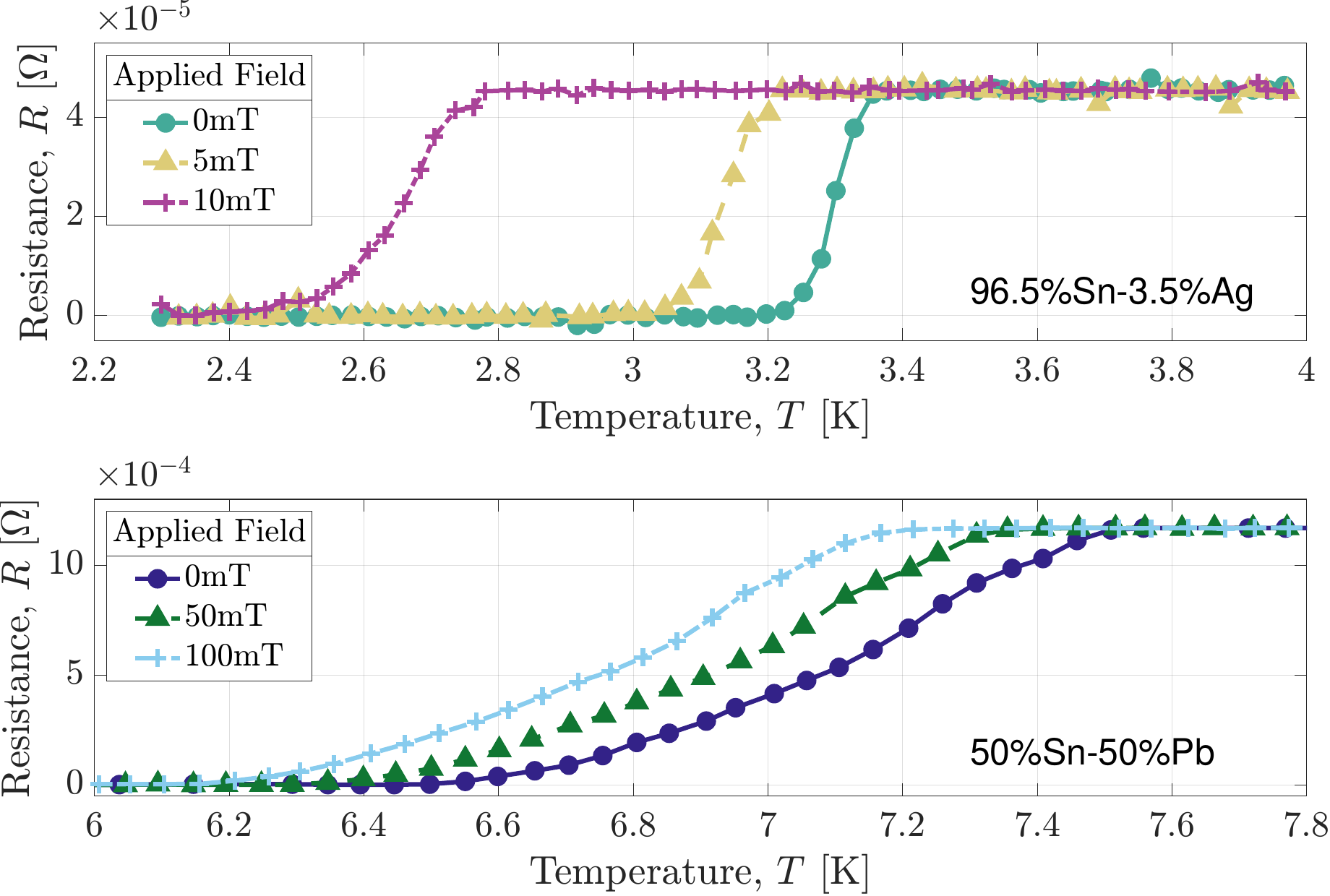}
\caption{DC resistance measurements versus temperature at different applied magnetic fields for the 96.5\%Sn-3.5\%Ag sample (top panel) and the 50\%Sn-50\%Pb sample (bottom panel).}
\label{fig:2}
\end{figure}

In Figure~\ref{fig:2}, we show the resistance versus temperature data for the 96.5\%Sn-3.5\%Ag sample at several different fields. For this data, we determine $T_{c,0} = (3.31 \pm 0.04) \,{\rm K}$, which aligns closely with a recent measurement of a similar 98\%Sn-2\%Ag solder by Ng et al. \cite{ng_investigation_2023} of $T_{c,0} = 3.32 \,{\rm K}$. A discussion of this result is found in the next section. Expanding on their work, we also measured $T_c$ at 5 and 10$\,{\rm mT}$ fields to extrapolate $B_{c,0} = 34 \pm 5\,{\rm mT}$. Figure~\ref{fig:2} also shows our data for the 50\%Sn-50\%Pb solder.

\begin{figure}[]
\centering 
\includegraphics[width=0.75\textwidth]{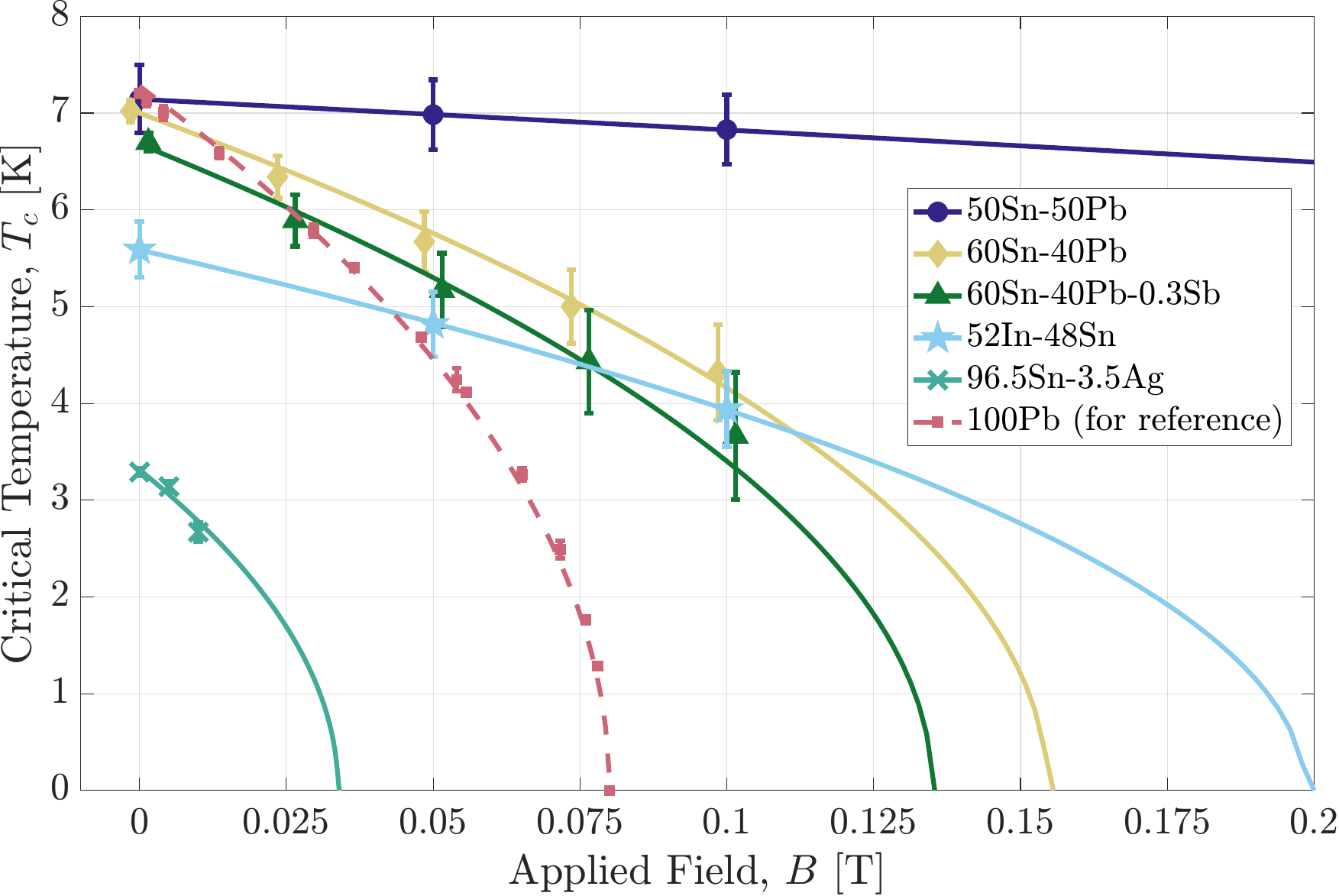}
\caption{The measured $T_c$ versus applied field $B$ for all our samples. Except for 96.5\%Sn-3.5\%Ag, all of our samples were measured at 0, 25, 50, 75, and 100$\,{\rm mT}$. Some data points (both doped and undoped 60\%Sn-40\%Pb) are offset horizontally to allow for easier visualization. The vertical error bars are from the measured transition width calculated by $\pm \,(90\% - 10\%\,{\rm resistance})/2$ (see text). The 100\%-pure Pb data from \textit{Decker et al. }\cite{decker_critical_1958} is included to provide a reference due to its well-known properties.
}
\label{fig:3}
\end{figure}

A plot and a table summarizing our measurements, the extracted parameters, and comparisons with literature values (where appropriate) for all the samples are shown in Figure~\ref{fig:3} and Table~\ref{tab:SC_properties}. We find that Equation~\ref{eq:WHH} fits our data well within the uncertainties. We were able to determine $T_{c,0}$ for our samples with precision between $\pm 1{\rm-}4\%$. The temperature and field range of our measurements limited the precision with which we could extract $B_{c,0}$. With the exception of 50\%Sn-50\%Pb, which was particularly poor due to its large $B_{c,0}$, we determined $B_{c,0}$ to within $\pm 7{\rm-}20\%$. Our uncertainty analysis, taking into account the transition width and data fitting uncertainties, contrasts with most values we found in existing literature, which largely ignored uncertainties. For example, the key reference of Warren and Bader \cite{warren_superconductivity_1969} that we compare several of our values with only reported uncertainty due to their temperature measurement.
\renewcommand{\thefootnote}{\alph{footnote}}

\begin{table*}[ht!]
\begin{minipage}{\textwidth}
    \centering
    \begin{tabular}{lccccc} \hline
 Solder& $T_{c,0}$ (K)& Literature $T_{c,0}$ (K)& $B_{c,0}$ (T)&Literature $B_{c,0}$ (T)&Reference\\ \hline
         50\%Sn-50\%Pb&  7.14 $\pm$ 0.32 &  $7.75 \pm 0.2$ \footnotemark[1] \footnotetext[1]{Uncertainties in Ref.~\cite{warren_superconductivity_1969} are from the temperature measurement uncertainty only. The temperature transition width at low fields was not described.} &  1.2 $\pm$ 1.8 & 0.21 $\pm$ 0.10 \footnotemark[2]\footnotetext[2]{Extrapolated from our fit with Eq.~\ref{eq:WHH} using the $B_c$ (T = 1.3 K) value from Ref.~\cite{warren_superconductivity_1969}.} &\cite{warren_superconductivity_1969}\\ 
         60\%Sn-40\%Pb& 7.00 $\pm$ 0.10 &  $7.05 \pm 0.2$ \footnotemark[1] & 0.155 $\pm$ 0.015 & 0.086 $\pm$ 0.015 \footnotemark[2]&\cite{warren_superconductivity_1969}\\ 
         60\%Sn-40\%Pb-0.3\%Sb& 6.68 $\pm$  0.09 & -- & 0.14 $\pm$ 0.01 & -- &\\
         52\%In-48\%Sn& 5.59 $\pm$ 0.26 & 4.5 to 7.5 \footnotemark[3] \footnotetext[3]{\emph{Levi et al.}\cite{levy_effect_1966} measured a first superconducting phase with $T_c = 4.5\,{\rm K}$ and a second superconducting phase of $5.5\,{\rm K}$. \emph{Tsui et al.}\cite{tsui_superconducting_2016} reported $T_c = 6.4\,{\rm K}$. \emph{Warren and Bader} \cite{warren_superconductivity_1969} reported $7.5\,{\rm K}$ for 50\%In-50\%Sn.}  & 0.20 $\pm$ 0.04 \footnotemark[4]\footnotetext[4]{Extrapolated from our fit with Eq.~(\ref{eq:WHH}). A measurement we made at $B=0.2\,{\rm T}$ did not exhibit a transition so value should be smaller.} & 0.34 & \cite{levy_effect_1966, tsui_superconducting_2016}\\
         96.5\%Sn-3.5\%Ag&  3.31 $\pm$ 0.04& 3.7 \footnotemark[5]\footnotetext[5]{From \emph{Pobell} \cite{Pobell1996} for 97\%Sn-3\%Ag. More recently, \textit{Ng et al.} \cite{ng_investigation_2023} reported $T_{c,0} = 3.32\,{\rm K}$ for a similar 98\%Sn-2\%Ag solder (the field dependence was not studied).} &  0.034 $\pm$ 0.005& -- & \\ \hline
    \end{tabular}
    \caption{Summary table of the extracted critical temperature and critical fields for all the solders studied.}
    \label{tab:SC_properties}
\end{minipage}
\end{table*}

\section{Discussion}\label{sec:Discussion}

Of our four candidates for AC magnet wiring, 50\%Sn-50\%Pb has the best properties, having a high critical temperature ($T_{c,0} = 7.14 \pm 0.32$ K) that does not drop significantly up to the measured 0.1$\,{\rm T}$. For 50\%Sn-50\%Pb, Warren and Bader \cite{warren_superconductivity_1969} has a measured value of $B_c(T=1.3\,{\rm K}) = 0.20 \pm 0.09\,{\rm T}$, suggesting that we expect $T_c$ to drop rapidly just beyond our measured field range. The discrepancy between the two values, assuming uncorrelated uncertainties, is $(0.61 \pm 0.38){\, \rm K}$. Although the manufacturing history of the solder mixture can have an effect on the measured $B_c$ \cite{livingston_superconducting_1967, chang_short_2020}, it should be noted that the properties of 60\%Sn-40\%Pb measured by both our groups agree very well. The $T_{c,0}$ value we report for 50\%Sn-50\%Pb is somewhat lower than by Warren and Bader \cite{warren_superconductivity_1969}, especially when considering the transition width (which was not described in the other study). The lower $T_c$ for 50\%Sn-50\%Pb in weak fields is an important consideration for using this material as superconducting current leads when direct contact with normal boiling liquid helium cannot be achieved.

For the 50\%Sn-50\%Pb solder, our best AC magnet wire candidate, we developed a technique for adding electrical insulation. We found that a mixture of 35\% 3M Scotchkote\textregistered ~Electrical Coating and 65\% acetone (by volume) produced a good viscosity for dip coating and offered a reasonable drying time. A wire spool was pulled through a bath of the coating solution in 30 minute steps to allow the coating to dry between. This was done twice. The coated wire was cryogenically cycled in liquid nitrogen and remained mechanically and electrically robust throughout this process.

For the 0.3\%-Sb-doped 60\%Sn-40\%Pb solder, we measured $T_{c,0} = 6.68 \pm 0.09 \,{\rm K}$, which was slightly less than that for non-doped 60\%Sn-40\%Pb. Higher levels of Sb doping in the Sn-Pb solder may have improved the superconducting properties. For example, \textit{Mousavi et al.} \cite{mousavi_novel_2016} found that 5\% Sb doping in Sn-In solders increased $T_c$ and $B_{c}$. We were not able to find the superconducting properties for Sb-doped 60\%Sn-40\%Pb elsewhere, despite its recommended use in \emph{Ekin} \cite{ekin_experimental_2006}.

For 52\%In-48\%Sn, we measured $T_{c,0} = (5.59 \pm 0.26)\,{\rm K}$. This falls within the wide range of values found in the literature (see the footnote). We did not see a second superconducting phase as reported by \textit{Levy et al.} \cite{levy_effect_1966}, but this is to be expected as we did not study the magnetic hysteresis curves of the eutectic mixture \cite{livingston_superconducting_1967}. The similar 50\%In-50\%Sn solder has a reported $T_{c,0} = 7.5\,{\rm K}$ \cite{warren_superconductivity_1969}, which is the value found in \emph{Ekin} \cite{ekin_experimental_2006}. In \emph{Pobell} \cite{Pobell1996}, the stated value for ``50-52\%In-50-48\%Sn'' is $T_{c,0} = 7.1 - 7.5\,{\rm K}$. For 52\%In-48\%Sn, we also report $B_{c,0} = (0.20 \pm 0.04)\,{\rm T}$ extracted by fitting with Eq.(\ref{eq:WHH}). We also performed a measurement at $B=0.2\,{\rm T}$ and did not see a transition; therefore, $B_{c,0}$ is likely below $0.2\,{\rm T}$. This does not agree with \textit{Tsui et al.} \cite{tsui_superconducting_2016}, who reported a value of $0.34\,{\rm T}$. In-Sn alloys may be particularly sensitive to sample history and measurement setup. A cautious approach is recommended if precise control over the superconducting transition of this solder is needed. For our application, despite the higher $B_c$ compared to 60\%Sn-40\%Pb (and about the same as 50\%Sn-50\%Pb), 52\%In-48\%Sn did not have a high $T_c$ compared to the other candidates, and therefore was not seriously considered.

For the 96.5\%Sn-3.5\%Ag solder, we obtained $T_{c,0} = (3.31 \pm 0.04\,{\rm K})$. The value in \emph{Pobell} \cite{Pobell1996} is $T_{c,0} = 3.7\,{\rm K}$ for 97\%Sn-3\%Ag, which is a significantly higher than our value. We note that the recent work of \textit{Ng et al.}\cite{ng_investigation_2023} measured $T_{c,0} = 3.32\,{\rm K}$ for a similar 98\%Sn-2\%Ag solder. There is a chance that the $T_{c,0} = 3.7\,{\rm K}$ found in \emph{Pobell} for 96.5\%Sn-3.5\%Ag is a mistake.\footnote{We are referring to Table 4.1 in \emph{Pobell} \cite{Pobell1996}. There is a chance the $T_{c,0} = 3.7\,{\rm K}$ value quoted is for 95.5\%Sn-3.5\%Ag-1.0\%Cd from \emph{Warren and Bader} \cite{warren_superconductivity_1969} (and not for 97\%Sn-3\%Ag). We could not fully confirm this because we were unable to find two of the seven the references (\emph{Meijer} (1976) and \emph{Gylling} (1971)) cited for the table due to their age.} This higher $T_{c,0} = 3.7\,{\rm K}$ value has been spread online, however.\footnote{Before the work of \emph{Ng et al.}\cite{ng_investigation_2023}, an online search yields a webpage from Meyer Tool \& Mfg. titled ``When Superconductivity is Unanticipated'' (https://www.mtm-inc.com/when-superconductivity-is-unanticipated.html. Retrieved Aug 9,2024), which gives 97\%Sn-3\%Ag as having $T_{c,0} = 3.7\,{\rm K}$.} For 96.5\%Sn-3.5\%Ag, we also measured $T_c$ at different fields and was able to determine $B_{c,0} = 0.034 \pm 0.005 \,{\rm T}$, which we did not find elsewhere. In addition to applications of low-silver tin solders for quantum computing flip chip assemblies \cite{ng_investigation_2023}, this ``soft silver solder'' has been recommended for use in cryogenic mechanical joints \cite{ekin_experimental_2006}. We have found this solder to be useful in regions in thermal contact with normal boiling liquid helium (4.2$\, {\rm K}$) where avoidance of the superconducting transition is important (e.g., for maintaining high magnetic field uniformity). We have successfully used this solder for cryogenic mechanical joints (copper to stainless steel, copper to brass, and stainless steel to brass), including for containing superfluid helium.

\section{Conclusion} \label{sec:conclusion}

We measured the superconducting transitions of five commercially available solders at low applied fields ($\leq 100\,{\rm mT}$), obtaining both $T_{c,0}$ and $B_{c,0}$ values. Throughout our analysis, we incorporated and propagated our measured 90\%-10\% transition widths following the IEC superconductivity recommendations. Given that $T_c$ of these solders are only a few degrees above $4\,{\rm K}$, and the wide adoption of dry refrigerators (or due to other engineering constraints) where magnet wiring or other components cannot be directly in contact with liquid helium, having the transition widths and ranges of $T_c$ is important when applying these materials in high-precision experiments.

For 60\%Sn-40\%Pb-0.3\%Sb, a recommended solder to reduce embrittlement, we report its superconducting properties for the first time. For 52\%In-48\%Sn, we report a $T_{c,0}$ lower than the typical range stated for this solder, such that care should be taken when using this material. For 96.5\%Sn-3.5\%Ag, we report a $T_{c,0}$ that is consistent with another recent study, but lower than the typical $T_{c,0}$ attributed to this solder. We report the $B_{c,0}$ of 96.5\%Sn-3.5\%Ag for the first time.

Our primary motivation was to determine the best wire material for our low-field $1\,{\rm kHz}$ superconducting AC magnets, which have stringent heating requirements and cannot be in contact with liquid helium. We confirmed that 50\%Sn-50\%Pb was the best candidate for this application, largely because it did not see a significant drop in $T_c$ as the magnetic field increased to 100$\,{\rm mT}$. We developed a method for insulating the 50\%Sn-50\%Pb wire so that it could be used to wind coils. Our measured $T_{c,0} = (7.14 \pm 0.32)\,{\rm K}$ for 50\%Sn-50\%Pb is somewhat lower than the only other available value that we found of $T_{c,0} = (7.75 \pm 0.2)\,{\rm K}$ from Warren and Bader \cite{warren_superconductivity_1969}.

\section*{Acknowledgements}{This work was supported in part by the US Department of Energy under grant number DE-FG02-97ER41042, the subcontract number 4000196635 administered through Oak Ridge National Laboratory and by the US National Science Foundation under Grant No. 2232117.}

%\appendix

%\section{Appendix title 1}
\bibliographystyle{elsarticle-num}
\bibliography{SCsolderBib.bib}
%\begin{thebibliography}{SCsolderBib.bib}
%\end{thebibliography}

\end{document}